\documentstyle[preprint,revtex]{aps}
\tightenlines
\begin{document}
\draft
\begin{title}
Combination Rules, Charge Symmetry,\\ and Hall Effect in Cuprates.
\end{title}
\author{A. J. Schofield\\ and\\ J. M. Wheatley}
\begin{instit}
Research Centre in Superconductivity, \\ University of Cambridge, \\
Madingley Road Cambridge, CB3 OHE, U.K.
\end{instit}
\begin{abstract}

The rule relating the observed Hall coefficient
to the spin and charge responses of the uniform doped Mott
insulator is
derived.
It is essential to
include the contribution of
holon and spinon three-current correlations to the effective
action of the  gauge field.
In the vicinity of the Mott insulating point the Hall coefficient
is holon dominated and weakly temperature dependent.
In the vicinity of a point of charge conjugation
symmetry the holon contribution to the
observed Hall coefficient is small: the Hall coefficient follows
the temperature dependence of the diamagnetic susceptibility with
a sign determined by the Fermi surface shape.

\end{abstract}

\narrowtext
The unusual temperature dependence of the Hall coefficient
of cuprate superconductors is a longstanding puzzle of the normal
state of these materials \cite{hall}. In YBa$_2$Cu$_3$O$_{6.92}$ for
instance,
the resistivity is
linear in
temperature while the inverse Hall angle follows
a $T^{2}$ law. This is not easily reconciled with a single
relaxation time, single band model where both
of these quantities are proportional to the same relaxation rate.
In other
respects the physical properties are those of a single correlated
band. It has been proposed that the observed behavior might be a
consequence of spin charge separation in a two dimensional
generalization of the Luttinger
liquid; the relaxation of spinons dominating the Hall
conductance, the relaxation of holons dominating the longitudinal
conductance \cite{pwahall}.

In this paper we employ an approach to the correlated electron fluid
provided by the uniform resonating valence bond state \cite{gauge}.
The symmetry of this state and a Fermi liquid state are the same,
and it contains a Fermi surface for spin excitations consistent
with Luttinger's theorem. The essential
distinction between the correlated state and a Fermi liquid state
is that the fermion effective mass is not required to diverge to
describe Mott insulating behavior. This feature arises from
the separation of spin and charge, with a constraint that the
spinon and holon currents in any physical state are equal
and opposite. Recently this model has been shown to give a
good account of the high temperature thermodynamics of the
t-J model \cite{eth}.
It also gives a qualitatively reasonable account of normal state
transport properties of a doped Mott insulator \cite{nl,ik}.
It is important to note, however, that the uniform model
apparently fails to reproduce the temperature
dependence of the Hall angle observed experimentally.

In this paper we investigate the question of the Hall
conductance in greater
detail.
The strategy employed is based
on the observation that $\sigma^{xy}$ derives from the
cubic term in the expansion of the action of the electromagnetic
field in the medium; in other words it is a
non-linear response. In the gauge model \cite{il}, the
action of the physical electromagnetic field
can be constructed systematically in terms of a small set of
parameters; namely, the susceptibilities and transport coefficients
of holons and spinons. Our aim is to do this {\it consistently} to
third order. The combination rule which relates the physical
Hall current to the spin and charge responses, and corrections
to the linear response are obtained in this way.

Spin and charge degrees of freedom make independent contributions
to the effective action of the gauge field.
Both species couple to the fictitious field $a$, which imposes the
constraint. Coupling to the physical
electromagnetic field $A$, with coupling strength, $e$,
is chosen to be via the bosons. With the choice of gauge
$a_0 = A_0 = 0$ the action is \cite{il}:
$S= \int_0^\beta d\tau \int d^2x {\cal L}$, the Lagrangean
density being
\begin{eqnarray}
{\cal L} = f^{\dag} \partial_\tau f  + b^{\dag} \partial_\tau b
- {1\over {2 m_f}} f^{\dag} \nabla^2 f - {1\over {2 m_b}} b^{\dag}
\nabla^2 b
\nonumber\\
- {\bf j}_f.a - {\bf j}_b.(a + eA)
\nonumber\\
+ {1\over {2 m_f}} \rho_f a^2
+ {1 \over {2 m_b}} \rho_b (a + e A)^2
\label{eq:lagrangian}
\end{eqnarray}
where $m$,$\rho$ and
$\bf j$
are effective masses, densities and currents respectively.
We omit the action of the free electromagnetic field.
Integration over
the particle fields yields
an effective action for coupled $a$ and $A$ fields:
\begin{eqnarray}
S[a,A] = {1\over 2}(a+eA)^2 \Pi_b^{(2)} + {1\over 2}
a^2 \Pi_f^{(2)} \nonumber\\+ {1\over 3!} (a+eA)^3 \Pi_b^{(3)} +
{1\over 3!} a^3 \Pi_f^{(3)}+\ldots
\label{eq:scubic}
\end{eqnarray}
$\Pi_{b(f)}^{(2)}$ and $\Pi_{b(f)}^{(3)}$ are the boson and
fermion two and three-current polarizations.
(The form of the three-current polarizations are shown in
Fig.\ \ref{fig2}.) We use a shorthand notation;
for instance $\Pi_f^{(3)}a^3$ represents $\sum_{p\omega,q\nu}
\Pi_f^{\alpha \beta \gamma}(p,\omega;q,\nu) a^\alpha_{p\omega}
a^\beta_{q\nu}
a^\gamma_{-p-q-\omega-\nu}$, with indices
$\alpha,\beta = \{x,y\}$ and Matsubara
frequencies $\omega,\nu= 2 \pi n /\beta$.
The physical current (equal to ${\bf j}_b$) flowing in a given
external field is the derivative of the effective action of the
electromagnetic field $S[A]$ with respect to $A$.
Since the Hall current is proportional
to $A^2$, we must retain terms to cubic
order in  Eq.~(\ref{eq:scubic}). Higher order non-linearities
not written explicitly are of course required for stability.

The $a$ field
integration is performed to get the effective action for the $A$
field alone. Before doing this, it is useful to extract the
contribution from the quadratic terms of  Eq.~(\ref{eq:scubic})
to $S[A]$. Making the change of variable
${\tilde a} = a + \Pi_b^{(2)} (\Pi_b^{(2)} +
\Pi_f^{(2)})^{-1}A$ and regrouping terms, we find:
\begin{eqnarray}
S[A]= {e^2\over 2} \Pi_b^{(2)} (\Pi_b^{(2)} +
 \Pi_f^{(2)})^{-1} \Pi_f^{(2)} A^2 \nonumber\\-{e^3 \over 3!}
[\Pi_b^{(3)} A_b^3 + \Pi_f^{(3)} A_f^3] - \log Z[A]
\label{eq:seff}
\end{eqnarray}
Here $A_b$ and $A_f$ are the `screened' external fields:
\FL
\begin{eqnarray}
A_b^\alpha(k\omega) &&= \Pi_{f}^{\alpha \beta}(k\omega)
(\Pi_{b}^{\beta \gamma}(k\omega) +
\Pi_{f}^{\beta \gamma}(k\omega))^{-1} A^\gamma(k\omega)\nonumber\\
A_f^\alpha(k\omega) &&= - \Pi_{b}^{\alpha \beta}(k\omega)
(\Pi_{b}^{\beta \gamma}(k\omega) +
\Pi_{f}^{\beta \gamma}(k\omega))^{-1} A^\gamma(k\omega)
\label{eq:fields}
\end{eqnarray}
and $Z[A]$ is:
\begin{eqnarray}
Z[A] = \int d {\tilde a}
\exp\left[ - {e^2\over 2}{\tilde a}(\Pi_b^{(3)}A_b^2 +
\Pi_f^{(3)}A_f^2)\right.
\nonumber\\
-{1\over 2} (\Pi_b^{(2)} + \Pi_f^{(2)} +
{e \over 2} (\Pi^{(3)}_b A_b + \Pi^{(3)}_f A_f) ) {\tilde a}^2
\nonumber\\
\left.-{1\over 3!} (\Pi_b^{(3)}+\Pi_f^{(3)}){\tilde a}^3
+ \ldots\right]
\label{eq:extrabit}
\end{eqnarray}
The derivation leading to  Eqs.~(\ref{eq:seff})-(\ref{eq:extrabit})
preserves the constraint at every step. As a check note that at
zero holon density the holon
polarizations vanish and that the screening factor appearing in
$A_f$ of
Eq.~(\ref{eq:fields})
is zero; thus $S[A]$ is independent of $A$ and no
physical currents flow in the presence of external fields in this
limit.

To expand $S[A]$ in powers of $A$, $Z[A]$ is computed in a saddle
point
approximation and a systematic expansion of
$\log Z[A]$ is made. The expansion to third order is
shown diagrammatically in Fig.\ \ref{fig3}. In the absence of external
fields, the
saddle point value $\langle \tilde a \rangle$ is zero; for finite
$A$ the shift is
$O(A^2)$. This shift in the saddle point contributes at third order
and
is responsible for diagrams of Fig.\ \ref{fig3}.d.

The symmetries of the underlying model Eq.~(\ref{eq:lagrangian})
constrain the form of the polarizations $\Pi$ and are essential
in analysing the diagrams
of Fig.\ \ref{fig3}. In addition to translations in space and
time and  permutation symmetry, the polarizations respect
inversion symmetry
$\Pi^{\alpha\beta \gamma} (k,\omega;p,\nu)=
-\Pi^{\alpha \beta \gamma}(-k,\omega;-p,\nu)$. Moreover, invariance
under static gauge transformations
$a(r,\tau) \rightarrow a(r,\tau) + \nabla f(r)$ implies that
$
\delta_{\omega,0} k^\alpha \Pi^{\alpha \beta }
(k\omega)=0
$
and
$
\delta_{\omega,0} k^\alpha \Pi^{\alpha \beta \gamma}
(k,\omega;p,\nu)=0
$ \cite{fuku}.
A special case is $\Pi^{\alpha \beta \gamma}
(k,\omega;-k,-\omega)= \Pi^{\alpha \gamma \beta}
(k,\omega;0,0) = 0 $, which guarantees that the effective action is
independent of $A^\alpha_{0,0}$.
Thus the first order diagram Fig.\ \ref{fig3}.a vanishes by gauge
symmetry.

The polarizability functions appearing in the expansion of
$\log Z[A]$ are
parameterized by responses of bosons and
fermions. For instance, for fermions,
$\Pi_f^{\alpha \beta}(q,\omega) =
(\delta^{\alpha \beta} -
{{q^\alpha q^\beta}/ q^2}) \pi^f_t(q\omega)
+({{q^\alpha q^\beta}/ q^2}) \pi^f_l (q,\omega)$,
where the transverse part
$\pi^f_t(q,\omega) = \chi_f q^2 +  \sigma_f |\omega |$ and the
longitudinal part
$\Pi^f_l(q,\omega) =
\nu_f
\omega^2 / q^2$. Here $\sigma_f$ is the conductivity, $\chi_f$
the orbital
response, and $\nu_f$ is the compressibility. Gauge invariance
implies that the expansion of
$\Pi_f^{\alpha \beta \gamma} ( k,\Omega; -k + q,-\Omega + \omega)$
begins at first order in $\omega$ and $q$. Thus for instance,
$\Pi_f^{xyy}(0,\Omega;q^x,-\Omega) = i \sigma_f^{xy} q^x \Omega$
where $\sigma_f^{xy}$ is the Hall response of spinons.

The quadratic diagram [3.(b)] gives the first corrections to the
combination rule (first term in Eq.~(\ref{eq:seff})) for the
conductivity and diamagnetic
susceptibility obtained at the gaussian level \cite{il}.
Since this diagram contains two
factors of $\Pi^{(3)}$ it yields terms proportional to $\omega^2$,
$\omega q$ or $q^2$; thus
the correction to the conductivity rule $\sigma^{-1} =
\sigma_b^{-1} + \sigma_f^{-1}$ vanishes. There is a finite
correction to the physical susceptibility ($\chi$) which can be
expressed as:
\begin{equation}
\chi^{-1} = {\chi_g^{-1} \over {1 + \Upsilon \chi_g^{-1} }}
\label{eq:susceptibility}
\end{equation}
where $ \chi_g^{-1}$ is the gaussian result
$\chi_b^{-1} + \chi_f^{-1}$ and $\Upsilon$ is
the coefficient of the $q^2$ term of diagram Fig.\ \ref{fig3}.b.
The low doping limit defined by
$\chi_b \ll \chi_f$, $\sigma_b \ll \sigma_f$,
$\sigma^{xy}_b \ll \sigma^{xy}_f$ is the regime with
marked non-Fermi liquid properties \cite{chib}.
Counting the
combination of
screening factors and boson polarizations shows that $\Upsilon$ is
proportional
to $\delta^2$. The correction is negligible close to the Mott
transition, above the
holon degeneracy temperature.

More precisely, in the low density limit the dominant contribution
to
$\Upsilon$ arises from the fermion polarizations:
\begin{eqnarray} \label{eq:diage}
\Upsilon &&\approx \left({\chi_b \over \chi_f}\right)^2
\beta \sum_\Omega
(\sigma^{xy}_f(\Omega) i \Omega )^2
{1\over 16 \pi} \int k dk (\pi^f_t)^{-2} \nonumber \\
&& \simeq - \left({\chi_b \over \chi_f}\right)^2 {1 \over 64 \pi^2}
\int_0^{\tau^{-1}} d\Omega { (\sigma^{xy}_f)^2
\over {\sigma \chi}}|\Omega| \nonumber \\
&&\sim - {{\sigma_f
\chi_b^2} \over \chi_f}
\end{eqnarray}

The Hall conductance corresponds to the second term of
Eq.~(\ref{eq:seff}) and
the saddle point contributions Fig.\ \ref{fig3}.c,d.
For the Hall geometry of
Fig.\ \ref{fig1} with small external momenta $q,0$
and $0,\omega$, diagram (c) gives a vanishing
contribution to the Hall current since it yields
terms proportional to $\omega^2 q$ or
$q^2 \omega$. Diagram (d) however makes a finite contribution
since it leads to
terms proportional to $i \omega q$. Dividing by the physical
conductivity we find
\FL
\begin{equation}
R_H  = R_{H}^b {\chi_f \over {\chi_f + \chi_b}} -
R_{H}^f {\chi_b \over {\chi_f + \chi_b}} + (R_{H}^b + R_{H}^f)
{\Lambda
\over {\chi_f + \chi_b}}
\label{eq:comb}
\end{equation}
where $\Lambda$ is the coefficient of the $q^2$ term of diagram
Fig.\ \ref{fig3}.e.
The first two terms in
Eq.~(\ref{eq:comb}) are identically the combination rule for the
Hall conductance
already used in the literature \cite{ik,mc}. In our convention,
$R_H^f$ is negative for a hole-like fermi surface.
The screening factors
express the fact that the
effective magnetic fields experienced by holons and spinons are
different and have a
simple heuristic derivation \cite{ik}.

In the low doping limit, inspection of diagram [3.(d)] shows that
$\Lambda$ is proportional to
$\delta$; we estimate (as above) $\Lambda \simeq - \sigma_f
\chi_b$.
Thus $R_H \simeq R_H^b$ since the boson contribution diverges as
$\delta^{-1}$.
Within Drude-Boltzmann theory, $R_H^b$ is weakly temperature
dependent and so Eq.~(\ref{eq:comb}) leads to a {\it temperature
independent Hall coefficient} at low doping \cite{ikw}.

The formalism developed above is valid beyond the low doping limit.
It is especially interesting to apply it close to a
point of charge symmetry in the lower Hubbard band. The significance
of this symmetry for
transport in strongly
correlated systems was first discussed by Beni and Coll \cite{beni}
who
studied the thermo-electric power
in the one dimensional Hubbard model. In common with the Hall
effect, the thermopower is odd under charge conjugation; it vanishes
in
the infinite U limit at the point of effective charge symmetry
(quarter filling).
Similarly, the three-current
polarization (and in particular $\sigma^{xy}$) vanishes identically
in a
charge symmetric system.

For half filling of hard core bosons on a lattice with
arbitrary off-site interaction $V_{ij}$, the model $H = \sum_{i,j}
t_{i,j} b^{\dag}_i b_j
- \mu \sum_i n_{i} + \sum_{i,j} V_{i,j} n_{i} n_{i}$ is invariant
under the
charge conjugation $b^{\dag}_i \rightarrow b_i$.
Half filling for
holons corresponds to quarter filling of the lower Hubbard band.
It is worth emphasising that a lattice effect underlies this
symmetry.

When $\Pi_b^{(3)} = 0$ the corrections to
the susceptibility Fig.\ \ref{fig3}.b and Hall effect
Fig.\ \ref{fig3}.d are simply related:
$\Upsilon = \Lambda \chi_b/(\chi_b + \chi_f)$.
Then, combining Eq.~(\ref{eq:comb})
and the expression for the diamagnetic susceptibilty
Eq.~(\ref{eq:susceptibility})
\begin{eqnarray}
{R_H \over \chi} &&= - {R_H^f \over \chi_b}
\left[ { {1 - {\Lambda \over \chi_b}} \over {1 + {\Lambda \over
\chi_f}}}\right]\nonumber \\
&& \sim  - { R_H^f \over \chi_f} [ 1 + O(\sigma_f)]
\label{eq:ccs}
\end{eqnarray}
where the second line assumes $\chi_b \ll \chi_f$ and estimating
$\Lambda \sim - \sigma_f \chi_b$ (as above).

According to  Eq.~(\ref{eq:ccs}) the {\it Hall coefficient follows the
temperature dependence of diamagnetic
susceptibility}. If charge symmetry occurs in the
non-Fermi liquid regime $\chi \sim T^{-1}$ then $R_H \sim T^{-1}$
or $\cot \Theta_H \sim T^2$ with a sign which is determined by
$R_H^f$. The possibility that
conjugation symmetry may be
present for charge but not
for spin is an elementary consequence of spin charge separation.
Eq.~(\ref{eq:ccs}) suggests that this can have dramatic
consequences in a Hall
effect experiment.

Crystalline YBa$_2$Cu$_3$O$_{6.92}$ displays the strongest
temperature dependence of $R_H$.
Susceptibilty
anisotropy measurements \cite{chiexp} on this material can be
interpreted in our units as a
diamagnetic
susceptibility  $\chi_d \sim \chi_f (100K/T)$ where
$\chi_f \sim 200K$ \cite{ik}.
If $R_H^f \sim -1$, then Eq.~(\ref{eq:ccs})
could account for the observed Hall coefficient. The observed small
magnitude of the thermopower is also consistent with a small holon
contribution.
However we caution that $\delta$ in this material
should correspond to about 0.2, which is far from quarter
filling ($\delta = 0.5$), the naive the point of charge symmetry.

To summarize, in this paper we have done two things. Firstly, we have
confirmed that
the spin charge separation present in the Ioffe-Larkin action
leads to $R_H \sim \delta^{-1}$ and weak temperature dependence
at low
doping. Secondly, we have pointed out that a temperature
dependence, which
agrees qualitatively
with experiment, occurs near to point of charge symmetry if the
system is in a non-Fermi liquid regime. This
anomalous temperature dependence is a direct consequence of spin
charge
separation as in the Luttinger liquid theory. Nevertheless the
predictions are
distinct from the latter which leads to strong temperature
dependence
independent of doping.

The authors thank J. Cooper and members of his group for
explaining the experimental situation, G. Gavazzi, T. Hsu,
B. Dou\c cot, D. Khmelnitskii and B. Farid for discussions and the
hospitality of the Institut Laue-Langevin where part of this work
was done.

\figure {In the presence of crossed electric
and magnetic fields, a transport current
$j^x_{\omega,0} = \sigma^{xx} i \omega A^x_{\omega,0}$,
a diamagnetic current
$j^y_{0,q_x} = \chi i q_x A^y_{0,q_x}$ and a
Hall current $j^y_{-\omega,-q_x}
= \omega q_x \sigma^{xy} A^x_{\omega,0}  A^y_{0,q_x}$ flow.
The Hall current is conjugate to the field $A^y_{-\omega,-q_x}$.
\label{fig1}}

\figure {Three-current polarizations: $\Pi^{\alpha \beta \gamma} =
\langle j^\alpha j^\beta j^\gamma \rangle - {1 \over m}
\delta^{\alpha \beta}
\langle \rho j^\gamma \rangle - {1 \over m} \delta^{\alpha \gamma}
\langle \rho j^\beta \rangle - {1 \over m} \delta^{\beta \gamma}
\langle \rho j^\alpha \rangle$.
The three-current correlation function is given by
diagrams (a), the density-current correlation functions
by diagram (b).\label{fig2}}

\figure {Expansion of the $\log Z[A]$ contribution to the
electromagnetic action.
Dashed lines indicate
the gauge field propagator  $(\Pi_b^{(2)} + \Pi_f^{(2)})^{-1}$.
Crosses accompanying external lines carrying the frequency and
momentum
of the external fields
indicate screening factors
$P_b = \Pi_f^{(2)}/(\Pi_f^{(2)} + \Pi_b^{(2)})$,
$P_f = -\Pi_b^{(2)}/(\Pi_f^{(2)} + \Pi_b^{(2)})$.
Three-current polarizations
are of four types: circles with no crosses indicate
$\Pi_b^{(3)} +\Pi_f^{(3)}$, with one cross
$ P_b \Pi_b^{(3)} + P_f \Pi_f^{(3)}$,
with two crosses
$ P_b^2 \Pi_b^{(3)} + P_f^2 \Pi_f^{(3)}$,
with three crosses
$ P_b^3 \Pi_b^{(3)} + P_f^3 \Pi_f^{(3)}$.
(a) Linear contribution vanishing by gauge symmetry.
(b) Quadratic term, giving a contribution
to the diamagnetic response for in-going frequency zero.
(c) Cubic diagram with vanishing
contribution to the hall current.
(d) Cubic term giving a finite correction to the Hall current.
(e) The correction factor $\Lambda$
computed in the text.\label{fig3}}


\begin{references}


\bibitem {hall} T.R. Chien, Z.Z. Wang and N.P. Ong, Phys. Rev.
Lett. {\bf 67} 15,
2088 (1991); T.R. Chien, D.A. Brawner, Z.Z. Wang and N.P. Ong,
Phys. Rev B
{\bf 43}, 6242 (1991); J. Clayhold, N.P. Ong, Z.Z. Wang,
J.M. Tarascon and P. Barboux,
Phys. Rev. B {\bf 39}, 7324 (1989); A. Carrington, A.P. Mackenzie,
C.T. Lin, C.T. Lin, J.R. Cooper, {\it preprint}.


\bibitem{pwahall} P.W. Anderson, Phys. Rev. Lett. {\bf 67},
2092 (1991).

\bibitem{gauge}
G.Baskaran  and P.W.Anderson, Phys. Rev. B {\bf 37}, 580 (1988).

\bibitem{eth} R. Hlubina, W. O. Puttika, T. M. Rice,
D. V. Khveshchenko, ETH {\it preprint}.

\bibitem{nl} Y.Nagaosa and P.A.Lee, Phys. Rev. Lett. {\bf 64},
2450 (1990).

\bibitem{ik} L.Ioffe and G.Kotliar, Phys.\ Rev.\ B {\bf  42}, 10348
(1990).

\bibitem{il} L.Ioffe and  A.Larkin, Phys. Rev. B {\bf 39}, 8988
(1989).

\bibitem{fuku} H. Fukuyama, H. Ebisawa, Y. Wada, Prog. Theor. Phys.
{\bf 42}, 494, (1969).

\bibitem{chib} The 2-D orbital susceptibilities of fermions and
bosons are
$\chi_{f}^{b}=\pm
\left[ \exp( \pm T_0/T) -1 \right]/(24 \pi m_{f}^{b})$.
For holons above their degeneracy temperature $T_0$ this becomes:
$T_0/(24 \pi m_b T)$.

\bibitem{mc} A very recent formal derivation of the first two terms
along the
above lines has been given: S. M. Manning and Y. Chen, submitted
to J. Phys. Cond. Matt.

\bibitem{ikw} L. B. Ioffe, V. Kalmeyer and P. B. Wiegmann,
Phys. Rev. B
{\bf 43}, 103483 (1991). These authors do not find a mechanism for
strong
temperature dependence of $R_H^b$ for $T > T_0$.

\bibitem{beni} G. Beni and C. Coll, Phys. Rev. B {\bf 11}, 573,
(1975).

\bibitem{chiexp} M. Miljak, G. Gollin, A. Hamzic and V. Zlatic,
Europhysics Letters, {\bf 9}, 723, (1989)



\end{references}
\end{document}